\documentclass{article}
\usepackage{ijcai18}
\usepackage{amsfonts}
\usepackage{times}
\usepackage{helvet}
\usepackage{courier}
\usepackage{url}
\usepackage{graphicx}
\usepackage{balance}
\usepackage{listings}
\usepackage{amsmath}
\usepackage{multirow}
\usepackage{adjustbox}
\usepackage{rotating}
\usepackage{color, colortbl}
\usepackage{subfig}
\usepackage{soul}
\usepackage{comment}
\usepackage{enumitem}
\usepackage{bbm}
\usepackage{tabularx}
\usepackage{tabu}
\usepackage{stmaryrd}
\usepackage{siunitx}
\newcolumntype{Y}{>{\centering\arraybackslash}X}

\definecolor{dkgreen}{rgb}{0,0.6,0}
\definecolor{Gray}{gray}{0.6}

\newcommand{\parabf}[1]{\noindent \textbf{#1}}

\newcommand{\OurApproach}{DeepRoad}

\newcommand{\Comment}[1]{}

\newcommand{\mmm}[1]{{\color{blue}{Mengshi:[#1]}}}
\newcommand{\lll}[1]{{\color{blue}{Lingming:[#1]}}}
\newcommand{\modelA}{Autumn}
\newcommand{\modelC}{Chauffeur}
\newcommand{\modelR}{Rwightman}

\newcommand{\nvidia}{NVIDIA}

\usepackage{caption}
\newcommand{\distance}{5pt}
\title{DeepRoad: GAN-based Metamorphic Autonomous Driving System Testing}

\author{
Mengshi Zhang$^1$, 
Yuqun Zhang$^2$\thanks{Corresponding author.}, 
Lingming Zhang$^3$,
Cong Liu$^3$, 
Sarfraz Khurshid$^1$
\\ 
$^1$ University of Texas at Austin\\
$^2$ Southern University of Science and Technology\\
$^3$ University of Texas at Dallas\\
mengshi.zhang@utexas.edu,
zhangyq@sustc.edu.cn,
lingming.zhang@utdallas.edu, \\
cong@utdallas.edu,
khurshid@ece.utexas.edu
}

\begin{document} 
\maketitle

\begin{abstract}
While Deep Neural Networks (DNNs) have established the fundamentals of DNN-based autonomous driving systems, they may exhibit erroneous behaviors and cause fatal accidents. To resolve the safety issues of autonomous driving systems, a recent set of testing techniques have been designed to automatically generate test cases, e.g., new input images transformed from the original ones. Unfortunately, many such generated input images often render inferior authenticity, lacking accurate semantic information of the driving scenes and hence compromising the resulting efficacy and reliability. 

In this paper, we propose DeepRoad, an unsupervised framework to automatically generate large amounts of accurate driving scenes to test the consistency of DNN-based autonomous driving systems across different scenes. In particular, DeepRoad delivers driving scenes with various weather conditions (including those with rather extreme conditions) by applying the Generative Adversarial Networks (GANs) along with the corresponding real-world weather scenes. Moreover, we have implemented DeepRoad to test three well-recognized DNN-based autonomous driving systems. Experimental results demonstrate that DeepRoad can detect thousands of behavioral inconsistencies for these systems.    
\end{abstract} 

\section{Introduction}

"\emph{The train came out of the long tunnel into the snow country. The earth lay white under the night sky. The train pulled up at a signal stop.}" 

The above quotation is from the first paragraph of fiction ``Snow Country'', which describes the scene when the protagonist Shimamura enters the snow country. Back to that time, train was the major vehicle for long-distance travels, while people have more choices today. Now, suppose Shimamura takes a Tesla in Autopliot mode~\cite{TeslaAutopilot}, after coming out of the tunnel, there raises a question: can the autopilot system operate safely on the snow-covered road, or the story just ends with a tragedy?

Autonomous driving is expected to transform the auto industry. Typically, autonomous driving refers to utilizing sensors (cameras, LiDAR, RADAR, GPS, etc) to automatically control vehicles without human intervention. The recent advances in Deep Neural Networks  (DNNs) enables autonomous driving systems to adapt their driving behaviors according to the dynamic environments. In particular, an end-to-end supervised learning framework is made possible to train a DNN for predicting driving behaviors (e.g., steering angles) by inputing driving scenes (e.g., images), using the $\langle$driving scene, driving behavior$\rangle$ pairs as the training data. For instance, DAVE-2~\cite{bojarski2016end}, released by \nvidia{} in 2016, can predict steering angles based on only driving scenes captured by a single front-centered camera of autonomous cars. 
%Their road tests claim that "for a typical drive in Monmouth County NJ from our office in Holmdel to Atlantic Highlands, we are autonomous approximately 98\% of the time."

%Though autonomous driving systems, such as DAVE-2, can approach impressive accuracy of determining driving behaviors, it is still possible that they make mistakes and lead to fatal accidents, e.g., Tesla's Autopilot-driven Model S slamming into the back of a firetruck [cite]. One major reason is DNNs are derived from finite datasets, which easily miss corner cases that result in bugs. On the other hand, to detect such erroneous driving behaviors, most state-of-the-art techniques demand manually collecting and labeling test data [cite] that significantly poses high costs.

Recent testing techniques~\cite{pei2017deepxplore,DeepTest} \Comment{\lll{there should be more work cited} \mmm{for the testing techniques, I only found these two, but for other target, there should be more papers},} demonstrate that adding error-inducing inputs to the training datasets can help improve the reliability and accuracy of existing autonomous driving models. 
%analogize the erroneous driving behaviors in DNN-based autonomous driving systems to the logical bugs in traditional software systems. Similar to detecting bugs and patching traditional software systems, they assume that detecting the erroneous driving behaviors of DNN-based autonomous driving systems and adding the error-inducing inputs to the training datasets can be helpful to improve  system reliability. Specifically, 
For example, the most recent DeepTest work~\cite{DeepTest} designs systematic ways to automatically generate test cases, seeking to mimic real-world driving scenes. Its main methodology is to transform training driving scenes by applying simple affine transformations and various effect filters such as blurring/fog/rain to the original image data, and then check if autonomous driving systems perform consistently among the original and transformed scenes. With large amounts of original and transformed driving scenes\Comment{derived metamorphic relations}, DeepTest can detect various erroneous inconsistent driving behaviors for some well-performed open-source autonomous driving models, in a cheap and quick manner. %in a much cheaper and quicker manner compared to the manual-collection-based techniques.

\begin{figure}[h]
\subfloat[]{\includegraphics[width=0.5\columnwidth]{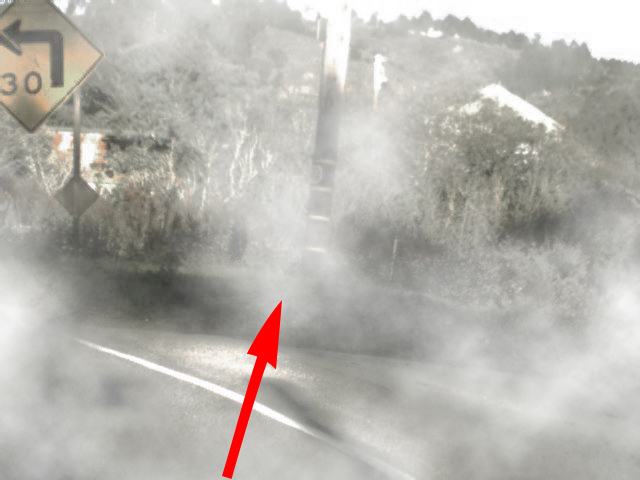} \label{fog}} 
\subfloat[]{\includegraphics[width=0.5\columnwidth]{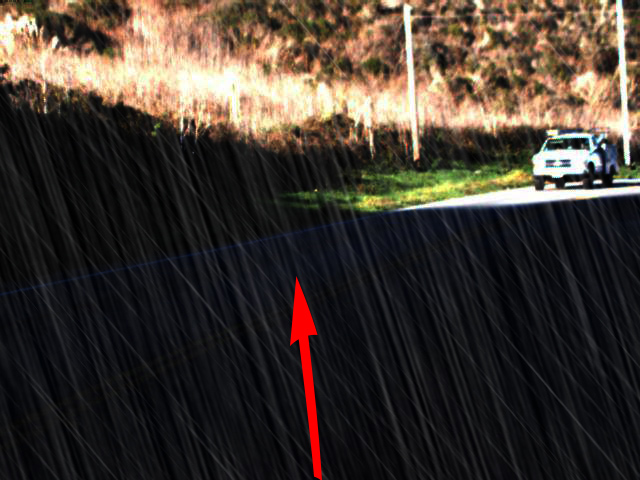} \label{rain}}
\centering
\caption{Foggy and rainy scenes via DeepTest}
\label{fig:deeptestEx}
\end{figure}

However, it is observed that the methodologies applied in DeepTest to generate test cases cannot accurately reflect the real-world driving scenes. Specifically, real-world driving scenes can rarely be affine-transformed and captured by the cameras of autonomous driving systems; the blurring/ fog/rain effects made by simply simulating the corresponding effects also appear to be unrealistic which compromises the efficacy and reliability of DeepTest. For instance, Figure~\ref{fog} shows the fog effect transformation applied in DeepTest\Comment{, where Figure~\ref{fog} is the original driving scene captured by the camera on an autonomous driving system and Figure~\ref{fog} is its transformed version via adding the fog effect}. It can be observed that Figure \ref{fog} is distorted.\Comment{\lll{we've put too many figs about deeptest, we should either remove the fog one or merge fog and rain by only keep the transformed pic.} \mmm{DeepTest figs has been merged, ref labels need to check consistent later.}} In particular, it appears to be synthesized by simply dimming the original image and mixing it with the scrambled ``smoke" effect. In addition, Figure~\ref{rain} shows the rain effect transformation applied in DeepTest. Similarly,\Comment{ when Figure~\ref{rain} applies the rain effect transformation on top of Figure~\ref{rain}, } DeepTest simply simulates rain by adding a group of lines over the original image. This rain effect transformation is even more distorted because usually when it rains, the camera tends to be wet and the image is highly possible to be blurred. \Comment{\lll{while we claim deepTest is not good, can we show some good figures from deepRoad?} \mmm{We present \OurApproach{} synthesized images in Figure~\ref{fig:DeepRoad-effect}.}\lll{the rain one is very unclear, can we have a clearer one? also why there is a "(a)" while there is only one sub-figure?} \mmm{Solved.}}
The fact that few test cases in DeepTest appear authentic to reflect the real-world driving scenes makes it difficult to determine whether the erroneous driving behaviors are caused by the flaws of the DNN-based models or the inadequacy of the testing technique itself.  Furthermore, there are many potential driving scenes that cannot be easily simulated with simple image processing. For instance, the snowy road condition requires different sophisticated transformations for the road and the roadside objects (such as trees).

% \begin{figure}[h]
% \includegraphics[width=1.0\columnwidth]{figure/synthetic_fig.png} 
% \centering
% \caption{Snow and rain effect in \OurApproach{}}
% \label{fig:DeepRoad-effect}
% \end{figure}

\Comment{Synthesizing distorted driving scenes by mixing effects with original driving scenes can be further argued as follows:
\begin{itemize}
\item 
\item It is highly likely that synthesizing large amounts of driving scenes by mixing effects with original driving scenes demands large amounts of manual operations using either drawing tools or Photoshop, which can be costly.\lll{this seems not true since photoshop can be applied in batch.} \mmm{The Photoshop is context-free, suppose the image content is changed, you have to tune the image manually. Phtotshop can do some simple operations in batch.}\lll{Of course, ps can, maybe we should just remove this point?} \mmm{I agree, we should discuss later.}
\end{itemize}
}

\begin{figure}[h]
\subfloat[]{\includegraphics[width=0.5\columnwidth]{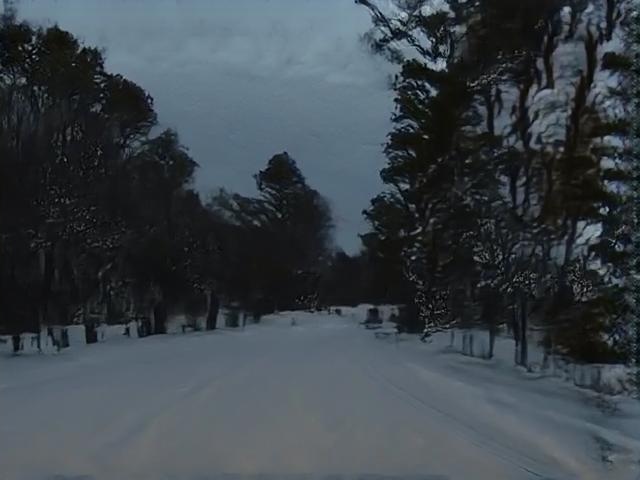} \label{syn_snow}} 
\subfloat[]{\includegraphics[width=0.5\columnwidth]{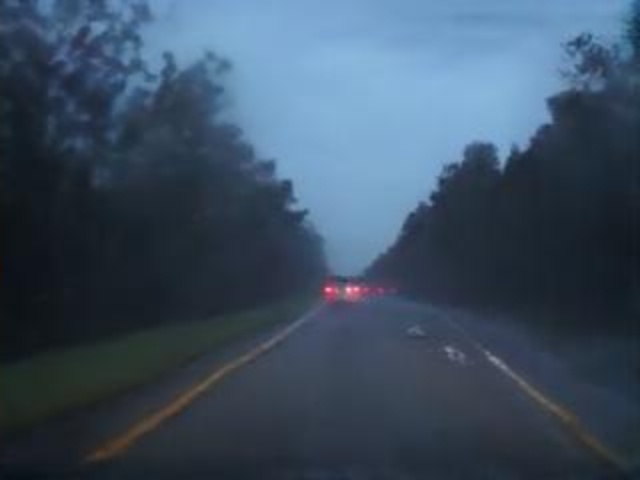} \label{syn_rain}}
\centering
\caption{Snowy and rainy scenes via \OurApproach{}}
\label{fig:deeproadEx}
\end{figure}

%\ccc{?Better to use two figures to show why existing testings that work on superficial, inaccurate driving scenes, fail in some real-world, but extreme weather scenarios, and then intuitively motivate the need of designing an automatic driving scene generator that is both accurate and scalable.?}

%\noindent\textbf{Our contributions.}
In order to automatically synthesize large amounts of authentic driving scenes for testing DNN-based autonomous driving systems, in this paper, we propose an unsupervised framework, namely \OurApproach{}, that employs a Generative Adversarial Network (GAN)-based technique~\cite{goodfellow2014generative} to deliver authentic driving scenes with various weather conditions which are rather difficult to be collected manually. Specifically, \OurApproach{} develops a metamorphic testing module for DNN-based autonomous systems, where the metamorphic relations are defined such that no matter how the driving scenes are synthesized to cope with different weather conditions, the driving behaviors are expected to be consistent with those under the corresponding original driving scenes. At this point, \OurApproach{} enables us to test the accuracy and reliability of existing DNN-based autonomous driving systems under different extreme weather scenarios, including heavy snow and hard rain, and can greatly complement the existing autonomous driving system testing approaches (such as DeepTest). For instance, 
Figure~\ref{fig:deeproadEx} presents the snowy and rainy scenes generated by DeepRoad (from fine scenes), which can hardly be distinguished from genuine ones and cannot be generated using simple transformations.

%We would like to know if the autonomous driving systems can perform correctly under some extreme scenes, such as snow-covered road or hard rain. The reason is we observe that typically, the image quality in these conditions are poor (specifically, the image becomes blurred or a lot of detailed information, such as lane marks are missing).

%However, these extreme scenes are not frequently happened, hence, collecting data from real word is not practical. On the other hand, manually editing/tuning images (PhotoShop) is expensive and not scalable. Therefore, the above limits inspires us to look for an automated method which can transfer a real driving image to one with extreme scenes.

%We choose GAN, a DNN method which can automatically train a generator, which can achieve our goal. When we get the generate image, we would like to investigate if the model can behavior consistently under different weather conditions to make sure the model stability.

Although our \OurApproach{} approach is general, and can be used to simulate various weather conditions, in this work, we first synthesize driving scenes under heavy snow and hard rain. In particular, based on the GAN technique, we collect images with the two extreme weather conditions from Youtube videos to transform real-world driving scenes and deliver them with the corresponding weather conditions. Subsequently, these synthesized driving scenes are used to test three well-recognized Udacity DNN-based autonomous driving systems~\cite{ModelCombo}. The experimental results reveal that \OurApproach{} can effectively detect thousands of behavioral inconsistencies of different levels for these systems, indicating a promising future for testing autonomous driving systems via GAN-based road scene transformation.    

The \textbf{contributions} of this paper are as follows.
\begin{itemize}
\item \textbf{Idea.} We propose the first GAN-based metamorphic testing approach, namely \OurApproach{}, to generate authentic driving scenes with various weather conditions for detecting autonomous driving system inconsistencies.
\item \textbf{Implementation.} We implement the proposed approach based on Pytorch and Python\Comment{\lll{which framework did you use? make it clear} \mmm{Pytorch, BTW, I used the authors' code and I am not sure if it is proper to claim this contribution, or if needed I can re-implement it after submission}} to synthesize driving scenes under heavy snow and hard rain based on training data collected from Youtube videos. 
\item \textbf{Evaluation.} We use \OurApproach{} to test well-recognized DNN-based autonomous driving models and successfully detect thousands of inconsistent driving behaviors for them. 
\end{itemize}

The rest of the paper is organized as follows. Section~\ref{sec2} introduces the background of autonomous driving systems and their existing testing techniques. Section~\ref{sec3} illustrates the overall approach of \OurApproach{}. Section~\ref{sec4} presents our experimental results on \OurApproach{}. Section~\ref{sec5} discusses some related work. Finally, Section~\ref{sec6} concludes this paper.\Comment{\lll{change all section numbers into refs} \mmm{Done.}}

\section{Background}
\label{sec2}

\begin{figure}[h]
\includegraphics[width=1.0\columnwidth]{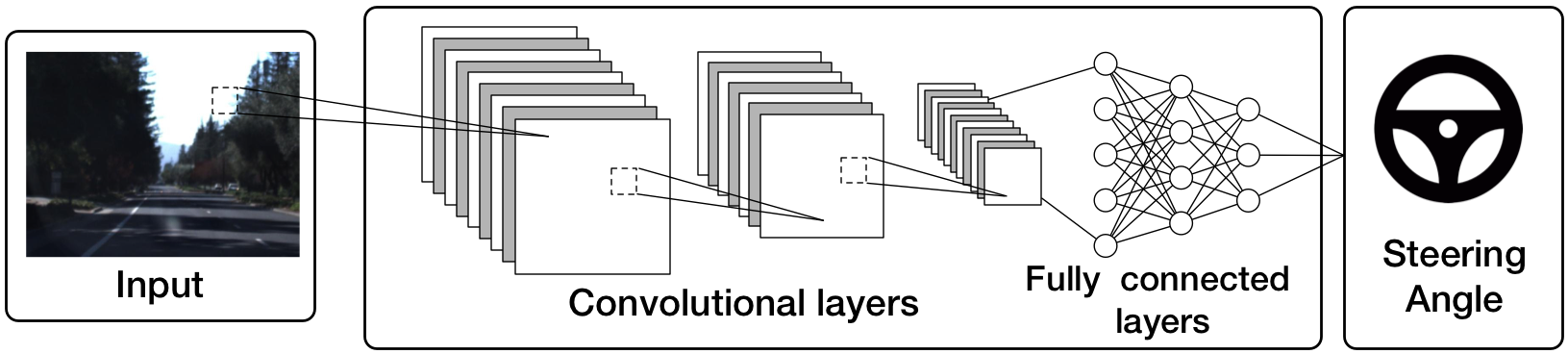}
\centering
\caption{Autonomous driving system on CNN}
\label{fig:cnn}
\end{figure}

%\lll{The first paragraph should be about DNN-based autonomous driving system.}
Nowadays, DNN-based autonomous driving systems have been rapidly evolving~\cite{Pomerlea,bojarski2016end}. For example, many major car manufacturers (including Tesla, GM, Volvo, Ford, BMV, Honda, and Daimler) and IT companies (including Waymo/Google, Uber, and Baidu) are working on building and testing various DNN-based autonomous driving systems. In DNN-based autonomous driving systems, the neural network models take the driving scenes captured by the sensors (LiDar, Radar, cameras, etc.) as input and output the driving behaviors (e.g., steering and braking control decisions).  In this work, we mainly focus on DNN-based autonomous driving systems with camera inputs and steering angle outputs. To date, feed-forward Convolutional Neural Network (CNN)~\cite{krizhevsky2012imagenet} and Recurrent Neural Network (RNN)~\cite{sak2014long} are the most widely used DNNs for autonomous driving systems. Figure~\ref{fig:cnn} shows an example CNN-based autonomous driving system. Shown in the figure, the system consists of an input (the camera image inputs) and an output layer (the steering angle), as well as multiple hidden layers. The use of convolution hidden layers allows weight sharing across multiple connections and can greatly save the training efforts; furthermore, its local-to-global recognition process actually coincides with the manual object recognition process.\Comment{\lll{can you enlarge the CNN one, and have camera input in the left and steering angle in the right? it's also important to mark what's conv layer and fully connected layers in the graph. we do not need to show RNN.}}

DNN-based autonomous driving systems are essentially software systems, which are error-prone and can lead to tragedies. For example, a Tesla Model S plowed into a fire truck at 65 mph while using Autopilot system~\cite{TeslaCrash}. To ensure the quality of software systems, many software testing techniques have been proposed in the literature~\cite{ammann2016introduction,mckeeman1998differential}, where typically, a set of specific test cases are generated to test if the software programs perform as expected. The process of determining whether the software performs as expected upon the given test inputs is known as the \emph{test oracle} problem~\cite{ammann2016introduction}. Despite the abundance of traditional software testing techniques, they cannot be directly applied for DNN-based systems since the logics of DNN-based softwares are learned from data with minimal human interference (like a blackbox) while the logics of traditional software programs are manually created. 

Recently, researchers have proposed various techniques to test DNN-based autonomous driving systems, e.g., DeepXplore~\cite{pei2017deepxplore} and DeepTest~\cite{DeepTest}. DeepXplore aims to automatically generate input images that can differentiate the behaviors of different DNN-based systems. However, it cannot be directly used to test one DNN-based autonomous driving system in isolation. The more recent DeepTest work utilizes some simple affine transformations and blurring/fog/rain effect filters to synthesize test cases to detect the inconsistent driving behaviors derived from the original and synthesized images.\Comment{ To our best knowledge, DeepTest is the first approach that intends to deliver methodologies for generating test cases for DNN-based autonomous driving systems. } Although DeepTest can be applied to test any DNN-based driving system, the synthesized images may be unrealistic, and it cannot simulate complex weather conditions (e.g., snowy scenes).

%\lll{The second paragraph introduces basic idea of testing, and the existing work on testing DNN-based autonomous driving system, e.g., DeepTest and DeepExplore and more...}
\newcommand{\MT}{MT}
\newcommand{\MR}{MR}
\newcommand{\prog}{p}
\newcommand{\iprog}{\hat{p}}
\newcommand{\itran}{f_I}
\newcommand{\otran}{f_O}
\newcommand{\DNN}{DNN}
\newcommand{\video}{\mathbb{I}}
\newcommand{\imageTrans}{\mathbb{T}}
\newcommand{\imageTran}{\tau}
\newcommand{\llb}{\llbracket}
\newcommand{\rrb}{\rrbracket}
\newcommand{\Lagr}{\mathcal{L}}

\section{Approach}
\label{sec3}
\subsection{Metamorphic DNN Testing}

Metamorphic Testing~\cite{metamorphic} (\MT) has been widely used to automatically generate tests to detect software bugs. The strength of \MT{} lies in its capability to automatically solve the test oracle problem via Metamorphic Relations (\MR). Formally, let $\prog{}$ be a program mathematical representation mapping program inputs into program outputs (e.g., $\prog\llb i\rrb=o$). Also, assuming $\itran$ and $\otran$ are two functions for transforming the input and output domain, respectively. Then, a \MR{} can be formed as:
\begin{equation}
\forall i.\; \prog\llb\itran(i)\rrb=\otran(\prog\llb i\rrb)
\end{equation}
With such \MR, we can test an actual implementation $\iprog$ of $\prog$ by checking whether $\iprog\llb\itran(i)\rrb=\otran(\iprog\llb i\rrb)$ for various inputs $i$. The idea of testing a program implementation via cross-checking inputs and outputs with \MR{} is called \MT. For instance, given a program implementing the $\sin$ function, we can use \MT{} to create various new tests without worrying about the test oracle problem. For any existing input $i$ for testing $\sin$, there are various facts that can directly serve as \MR{}, e.g., $\sin(-i)=-\sin(i)$ and $\sin(i+2\pi)=\sin(i)$. Note that $\itran(i)=\otran(i)=-i$ for the first example \MR, while $\itran(i)=i+2\pi\wedge \otran(i)=i$ for the second. With such \MR{}s, we can transform the existing test inputs according to $\itran$ to generate additional tests, and check the output based on $\otran$. 

In this work, we further apply \MT{} to test DNN-based autonomous driving systems. Formally, let $\DNN$ be a DNN-based autonomous driving system that continuously maps each image into predicted steering angle signal (e.g., turn left for 15$^{\circ}$). Then, given the original image stream $\video$, we can define various image transformations $\imageTrans$ that simply change the road scene (detailed shown in Section~\ref{sec:gan}) and do not impact the prediction results for each image $i\in \video$ (e.g., the predicted direction should be the approximately the same for the same road condition during fine and rainy days). In this way, we have the following \MR{} to test DNN with additional transformed inputs:
\begin{equation}
\forall i\in\video\wedge \forall \imageTran\in\imageTrans.\; \DNN\llb \imageTran(i)\rrb=\DNN\llb i\rrb
\end{equation}

\subsection{DNN-based Road Scene Transformation}
\label{sec:gan}
The recent work DeepTest~\cite{DeepTest} also applied \MT{} to test DNN-based autopilot systems. However, it only performs simple synthetic image transformation, such as adding simple blurring/fog/rain effect filters, and thus has the following limitations: (1) DeepTest may generate unrealistic images (e.g., the rainy scene shown in Figure~\ref{rain}\Comment{\lll{add some distorted images from DeepTest} \mmm{Are current figures good?}}), (2) DeepTest cannot simulate complex road scene transformations (e.g., snowy scenes).

To complement DeepTest and generate various real-world road scenes fully automatically, in this work, we leverage UNIT~\cite{liu2017unsupervised}, a recent published DNN-based method to perform unsupervised image-to-image transformation. One insight of UNIT is a paired images in different domains can be projected into a shared-latent space and have the same latent representation. In this way, given a new image from one domain (e.g., the original driving scene), UNIT can automatically generate its corresponding version in the other domain (e.g., rainy driving scene). Overall, UNIT is composed by generative adversarial networks (GANs)~\cite{goodfellow2014generative} and variational autoencoders (VAEs)~\cite{kingma2013auto}. 

\begin{figure}[h]
\includegraphics[width=0.7\columnwidth]{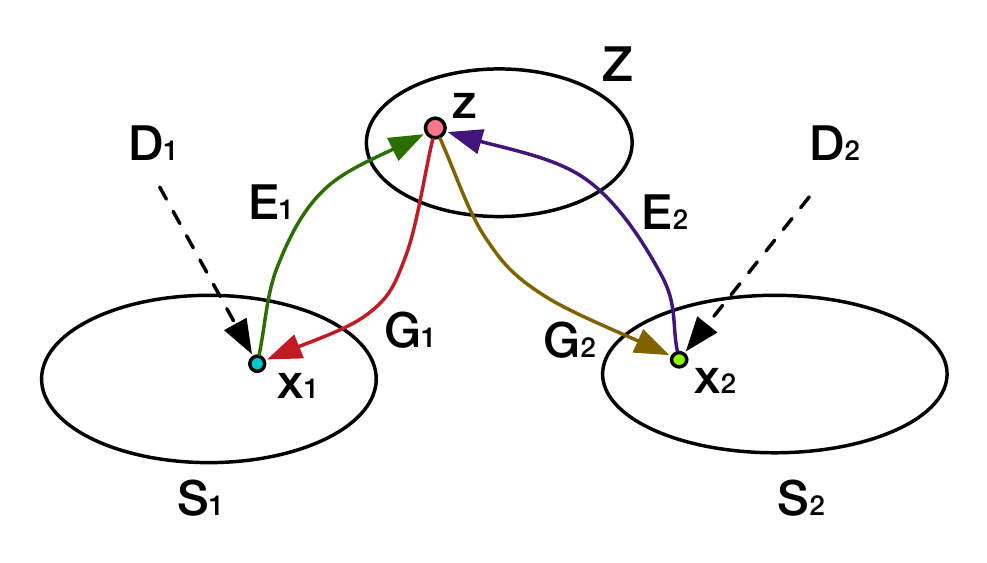} 
\centering
\caption{Structure of UNIT}
\label{fig:unit-structure}
\end{figure}

Figure~\ref{fig:unit-structure} presents the basic structure of UNIT, $S_{1}$ and $S_{2}$ denote two different domains (e.g., images include fine and rainy scenes, respectively), $E_{1}$ and $E_{2}$ are two autoencoders which can project the images from $S_{1}$ and $S_{2}$ to the shared-latent space $Z$. Suppose $x_{1}$ and $x_{2}$ are corresponding images which share the same contents, ideally, $E_{1}$ and $E_{2}$ would encode them to the same latent vector $z$. $G_{1}$ and $G_{2}$ are two domain specific generators which can translate a latent vector back to $S_{1}$ and $S_{2}$, respectively. $D_{1}$ and $D_{2}$ are two discriminators which can detect whether the image belongs $S_{1}$ and $S_{2}$, respectively. Ideally, the discriminators cannot differentiate whether the input image is from the target domain or a well-trained generator. Based on the autoencoders and generators, UNIT can be used to transform images between two domains. For instance, image $x_1$ can be transformed to $S_{2}$ by $G_{2}(E_{1}(x_{1}))$.

% input images with wiper swiping

The learning objective of UNIT can be decomposed to optimize three costs:
\begin{itemize}
  \item \textbf{VAE loss} minimizes the loss of the image reconstruction for each $\langle E_{i}, G_{i}\rangle$ pair. 
  \item \textbf{GAN loss} achieves the equilibrium point in the minimax game for each $\langle G_{i}, D{i}\rangle$, where $D_{i}$ aims to discriminate between images from the domain distribution and candidates produced by $G_{i}$ aiming to fool $D_{i}$.
  \item \textbf{Cycle-consistency loss} minimizes the loss of cycle-reconstruction for each $\langle E{i}, G{j}, E_{j}, G{i}\rangle$, ideally, $x_{1} = G_{1}(E_{2}(G_{2}(E_{1}(x_{1}))))$ and $x_{2} = G_{2}(E_{1}(G_{1}(E_{2}(x_{2}))))$
\end{itemize}

The total loss can be summarized as follows:
\begin{align*}
\min_{E_{1},E_{2},G_{1},G_{2}}\max_{D_{1},D_{2}} \Lagr_{CC_{1}}(E_{1},G_{2}, E_{2}, G_{1}) \\
 + \Lagr_{CC_{2}}(E_{2},G_{1}, E_{1}, G_{2}) \\
 + \Lagr_{VAE_{1}}(E_{1}, G_{1}) + \Lagr_{VAE_{2}}(E_{2}, G_{2}) \\
 + \Lagr_{GAN_{1}}(D_{1}, G_{1}) + \Lagr_{GAN_{2}}(D_{2}, G_{2})
%  + \Lagr_{CC_{1}}(E_{1},G_{2}, E_{2}, G_{1}) \\
%  + 
\end{align*}

\Comment{\lll{why min is only related to e1,e2, g1? for example, for VAE, i believe you also need to do min for g2.}
\mmm{actually it is minmax (a whole word), however I dont know how to write it}
\lll{can you also list separate loss func for each of the three ones?}}

\subsection{Overall Framework}
% real image from Youtube
\begin{figure}[h]
\includegraphics[width=0.9\columnwidth]{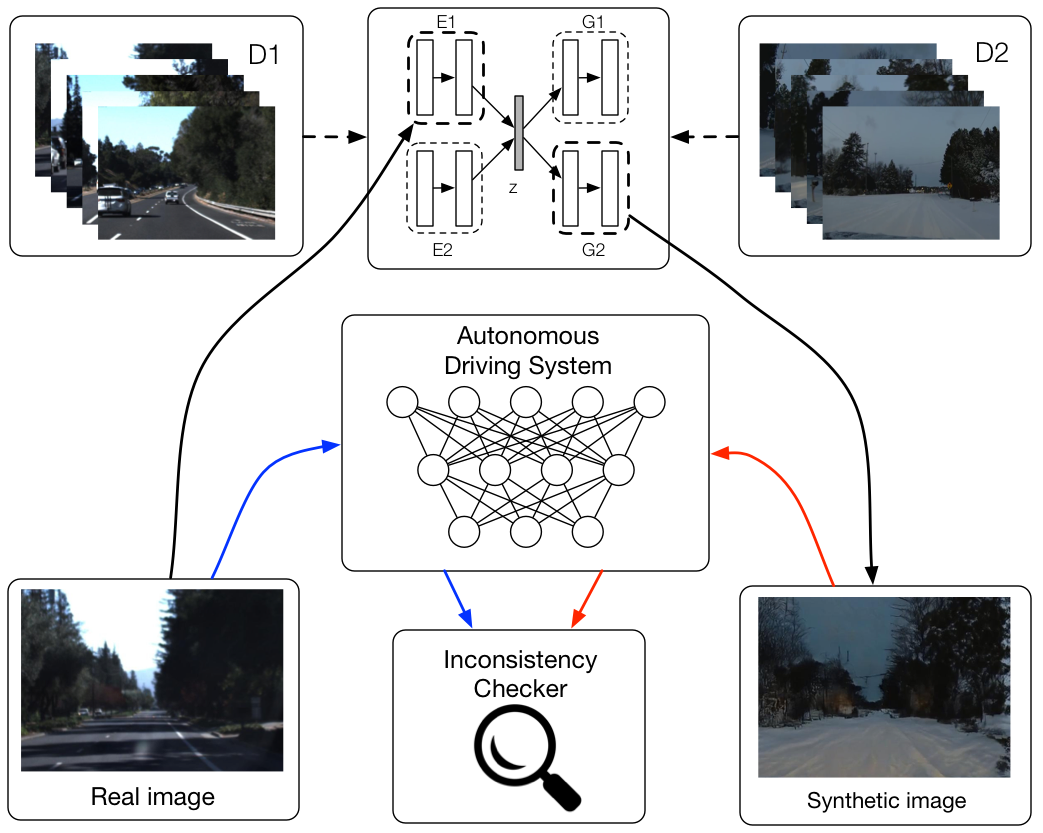}
\centering
\caption{Framework of \OurApproach{}}
\label{fig:framework}
\end{figure}

Figure~\ref{fig:framework} shows the overall design of our metamorphic testing framework for DNN-based autonomous driving systems, DeepRoad. Shown in the figure, DeepRoad firstly takes unpaired training images from two target domains (e.g., one fine driving scene dataset and one rainy driving scene dataset), and utilizes the unsupervised UNIT to map the two scene domains to the same latent space using the loss functions presented in Section~\ref{sec:gan}. In this work, we sample images from the real-world Udacity Challenge 2 dataset~\cite{UdacityDataset} (fine scenes) and Youtube video (snowy or rainy scenes~\cite{YoutubeRain,YoutubeSnow}) and feed them into UNIT for training. After it is well-trained, DeepRoad uses the UNIT model to transform the whole real-world Udacity driving dataset to another scene (e.g., snowy or rainy scenes). That is, given any original fine driving scene $i$, DeepRoad can apply the trained UNIT model to derive its corresponding version in another weather condition (e.g., rainy scene), $\imageTran(i)$. Then, DeepRoad will feed each pair of real and synthesized images to the autonomous driving systems under test (i.e., $DNN$), and compare their prediction results (i.e., $\DNN\llb \imageTran(i)\rrb ?= \DNN\llb i\rrb$) to detect any inconsistent behaviors.
Since the road scenes should not largely impact the steering angles, any inconsistency may indicate correctness or robustness issues of the systems under test. Note that although in this work we only explore the rainy and snowy scene transformations, our DeepRoad approach is general, can support any scene transformation supported by the underlying UNIT model.

\Comment{
\lll{you can make the autonomous driving system fig larger or wider in the framework fig}
\mmm{solved.}
}

\section{Experiments}
\label{sec4}

\subsection{Data}
We use a real-world dataset released by Udacity~\cite{Udacity} as a baseline to check the inconsistency of autonomous driving systems. From the dataset, we select two episodes of high-way driving video where obvious changes of lighting and road conditions can be observed among frames. To train our UNIT model, we also collect images of extreme scenarios from Youtube. In the experiments, we select snow and hard rain, two extreme weather conditions to transform real-world driving images. To make the variance of collected images relatively large, we only search for videos which is longer than 20mins. In the scenario of hard rain, the video would record wipers swiping windows, and in the data preprocessing phase, we manually check and filter out those images. Note that all images used in the experiments are cropped and resized to $240 \times 320$, and we have performed down-sampling for Youtube videos to skip consecutive frames with close contents. The detailed information is present in Table~\ref{tab:img_details}. \Comment{\lll{we only use one video for rain and one for snow? why 0.5 hour video only generates 1000 frames?} \mmm{Actually we can extract a lot of frames, however, the difference between consecutive frames are very small, and I down-sampled them, moreover, some frames which contain wiper are also filtered.}}
\begin{table}
\centering
\caption{\label{tab:img_details}Details of image sets}

\begin{tabular}{|c|c|c|c|} 
\hline
\bf{Dataset} & \bf{Frame} & \bf{Duration} & \bf{Weather Cond.}\\  \hline \hline
Udacity Ep1 & 15212 & N.A. & Sunshine  \\ \hline
Udacity Ep2 & 5614 & N.A. & Sunshine  \\ \hline
Youtube Ep1 & 1000 & 28:55 & Heavy snow  \\ \hline
Youtube Ep2 & 1000 & 1:09:03 & Hard rain  \\ \hline
\end{tabular}

\end{table} 
% Rain: https://www.youtube.com/watch?v=O88fXBx-Qdg
% Snow: https://www.youtube.com/watch?v=ps56tnQG8V0

% real image from Youtube
\begin{figure*}[h]
\includegraphics[width=2\columnwidth]{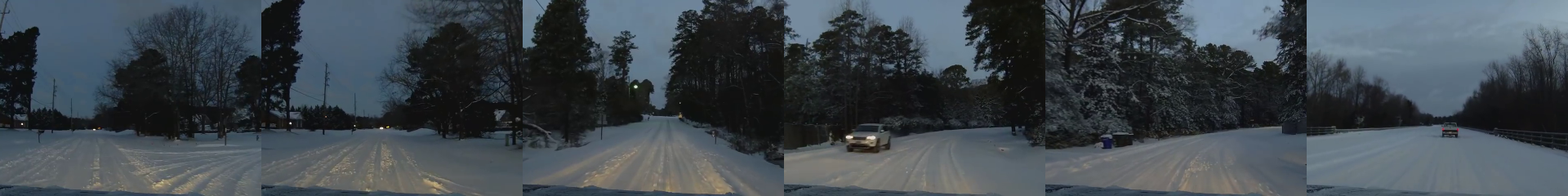}
\includegraphics[width=2\columnwidth]{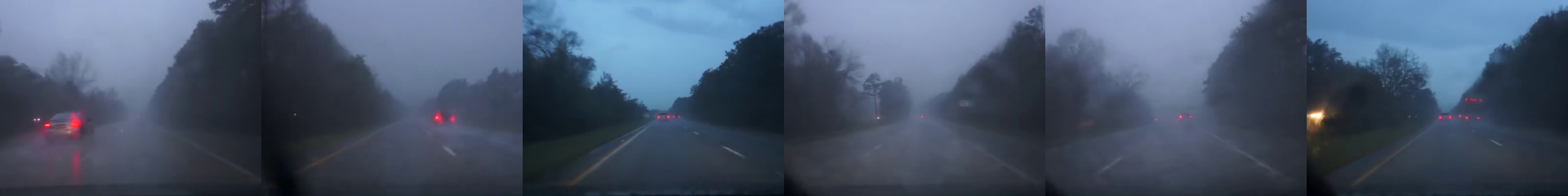} 
\centering
\caption{Images collected from Youtube}
\label{fig:youtube}
%\end{figure*}
% GAN-generated image
%\begin{figure*}[h]
\includegraphics[width=2\columnwidth]{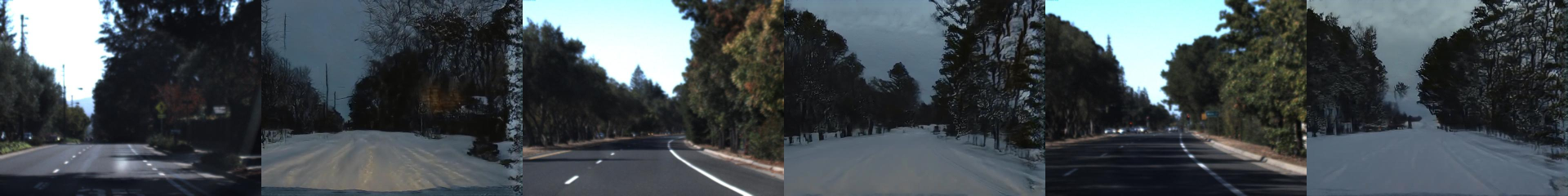}
\includegraphics[width=2\columnwidth]{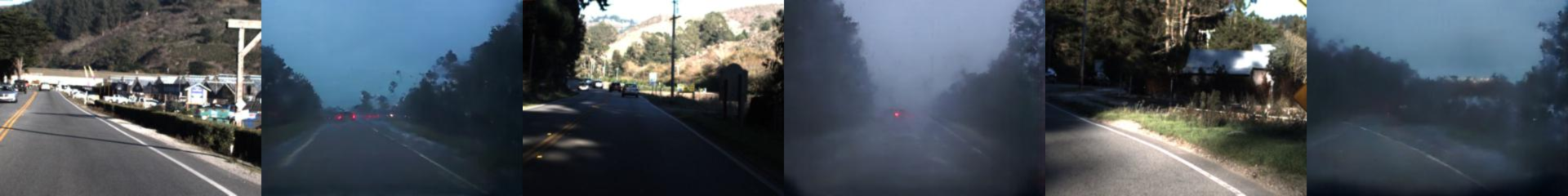} 
\centering
\caption{Real and GAN-generated images.}
\label{fig:real_fake_img}
\end{figure*}

\subsection{Models}
We evaluate our framework on three DNN-based autonomous driving models which are released by Udacity~\cite{Udacity}: \texttt{\modelA{}}~\cite {Autumn}, \texttt{\modelC{}}~\cite{Chauffeur}, and \texttt{\modelR{}}~\cite{Rwightman}. We choose these three models as their pre-trained model are public and can be evaluated directly on the synthesized datasets. To be specific, the model details of \texttt{\modelR{}} are not publicly released, however, just like black-box testing, our approach aims to detect the inconsistencies of the model instead of localizing software faults, hence, we still use \texttt{\modelR{}} for the evaluation. 

\parabf{\modelA{}.} \texttt{\modelA{}} is composed by a data preprocessing module and a CNN. Specifically, \texttt{\modelA{}} first computes the optical flow of input images and input them to a CNN to predict the steering angles. The architecture of \texttt{\modelA{}} is: three 5x5 conv layers with stride 2 pluses two 3x3 conv layers and followed by five fully-connected layers with dropout. The model is implemented by OpenCV, Tensorflow and Keras. 

\parabf{\modelC{}.} \texttt{\modelC{}} consists of one CNN and one RNN module with LSTM. The work flow is that CNN firstly extracts the features of input images and then utilizes RNN to predict the steering angle from previous 100 consecutive images. This model is also implemented by Tensorflow and Keras.

% Model image
\begin{figure*}[h]
\includegraphics[width=2\columnwidth]{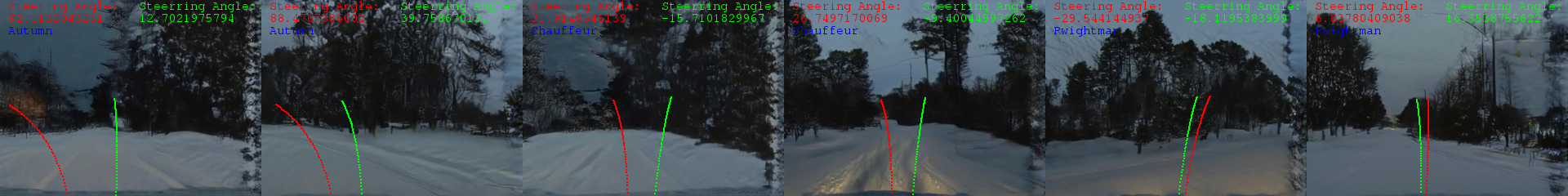}
\includegraphics[width=2\columnwidth]{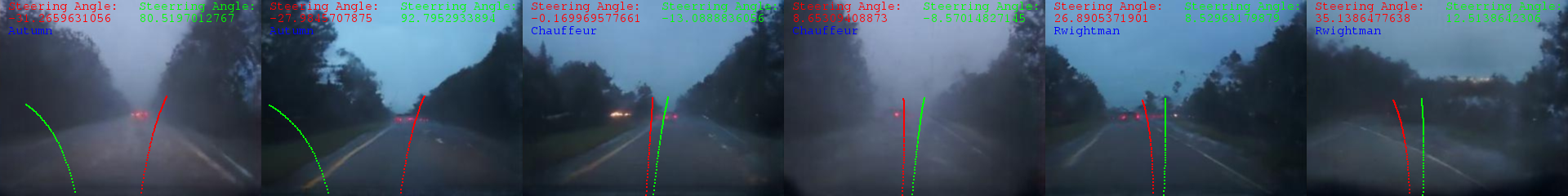}
\centering
\caption{Inconsistency of steering angle prediction on real and synthesized images.}
\label{fig:result_img}
\end{figure*}

\subsection{Metric}
Based on our assumption, an autonomous driving system is consistent if its steering angle prediction does not change after modifying the weather condition of driving images. However, this assumption is too strong to be practical since minor steering angle change incurred by the scene change may still fall into the safe zone\Comment{the transformation cannot keep all semantic information such as cars and buildings}. Hence, similar with prior work~\cite{DeepTest}, we relax the assumption and accept the prediction if the difference between the predicted steering angles of original and transformed images can be within an error bound. We define the number of inconsistent behaviors of autonomous driving systems as follows:\[
IB(\DNN, \video) = \sum_{i\in\video}{f(|\DNN\llb i\rrb - \DNN\llb \imageTran(i)\rrb)| > \epsilon)}
\],
where $\DNN$ denotes the autonomous driving model and $\video$ is the real-world driving dataset. $i$ denotes the $i$th image in $\video$. $\imageTran$ denotes the image generator/transformer which can change the weather condition of the input image. Function $f$ outputs 1 or 0 if and only if the input is $True$ or $False$ and $\epsilon$ is the error bound.

\subsection{Results}
\parabf{Quality of generated images}
We first present several Youtube frames as ground truth in Figure~\ref{fig:youtube} to help readers check the quality of generated images. In Figure~\ref{fig:real_fake_img}, we list real and GAN-generated images pairs, where the two rows present the transformation of Udacity dataset to snowy and rainy scenes, respectively, and the odd and even columns present original and GAN-generated images, respectively. Qualitatively, the GAN-generated images are visually similar to the images collected from Youtube and they also can keep the major semantic information (such as the shape of tree and road) of the original images. Interestingly, in the first snowy image in Figure~\ref{fig:real_fake_img}, the sky is relatively dark and GAN can successfully render the snow texture and the light in front of the car. In the second column, the sharpness of rainy images are relatively low and this is consistent to the real scene showed in Figure~\ref{fig:youtube}. Our results are consistent with the original UNIT work~\cite{liu2017unsupervised}, and further demonstrate the effectiveness of UNIT for image transformation.\Comment{\lll{it seems the first two figures in the second row of figure 8 do not match; check if there is any bug} \mmm{I have check these figures,  they are matched although the background is modified.}
\lll{we should also add some sentence to show that this further confirms the effectiveness of the UNIT approach to give other people credit}}

\parabf{Inconsistency of autonomous driving models}
We further present examples for the detected inconsistent autonomous driving behaviors in Figure~\ref{fig:result_img}. In the figure, each row shows the scenes of snow and rain, respectively. In each sub-figure, the blue caption indicates the model name, while the red and green captions indicate the predicted steering angles on the real and synthesized images, respectively. The curves visualize the predictions which help check the differences. \Comment{Note that actual steering curves may be different to presented ones\lll{why?}.
\mmm[This curves are generated by some parameters and these configurations are different on different cars. This is just a tiny issue and we can remove it if it is confused.]} From the figure we can observe that model \texttt{\modelA} (the first two columns) has the highest inconsistency number on both scenes; in contrast, model \texttt{\modelR} (the last two columns) is the most stable model under different scenes. This figure shows that \OurApproach{} is able to find inconsistent behaviors under different road scenes for real-world autonomous driving systems. For example, a model like \texttt{\modelA} or \texttt{\modelC}~\cite{UdRank} (they are both ranked higher than \texttt{\modelR} in the Udacity challenge) may work perfectly in a fine day\Comment{\lll{can we have some data to show that \texttt{\modelA} performs well on fine days }, \mmm{I use \modelC to show our effectiveness. Solved}} but can crash into the curbside (or even worse, the oncoming cars) in a rainy or snowy day (shown in Figure~\ref{fig:result_img}).

Table~\ref{tab:inconsistncy_behavior} presents the detailed number of detected inconsistent behaviors under different weather conditions and error bounds for each studied autonomous driving model on the Udacity dataset. For example, when using the error bound of \ang{10} and the rainy scenes, \OurApproach{} detects 5279, 710, and 656 inconsistent behaviors for \texttt{\modelA}, \texttt{\modelC}, and \texttt{\modelR}, respectively. From the table we can observe that the inconsistency number of \texttt{\modelA} is the highest under both weather conditions. We think one potential reason is that \texttt{\modelA} is purely based on CNN, and does not utilized prior history information (e.g., via RNN), and thus may not always perform well in all road scenes. On the other hand, \texttt{\modelR} performs the most consistently than the other two models under all error bounds.\Comment{ However, the prediction of \texttt{\modelC} is more stable in the hard-rain scene \mmm{when error bound is $\ang{10}$}, where \texttt{\modelR} performs reversely \lll{this sentence is not right}.}\Comment{ Table~\ref{tab:inconsistncy_behavior} further presents the number of inconsistent behaviors of the 3 studied autonomous driving systems detected by \OurApproach{} on the real-world Udacity dataset. Overall, the inconsistent behavior information conforms with the inconsistent numbers shown in Table~\ref{tab:inconsistncy}.} This result presents a very interesting phenomenon -- \OurApproach{} can not only detect thousands of inconsistent behaviors of the studied autonomous driving systems, but can also measure different autonomous systems in terms of their robustness. For example, with the original Udacity dataset, it is hard to find the limitations of autonomous driving systems like \texttt{\modelA}.

\Comment{the scene sensitivity is different for different models and \OurApproach{} is effective to detect their difference. }

\begin{table}
\centering
\caption{\label{tab:inconsistncy_behavior}Number of inconsistency behavior of three models under different weather conditions}

\begin{tabular}{|c|c|c|c|c|c|} 
\hline
\bf{} & \multirow{2}{*}{\bf{Model}} & \multicolumn{4}{c|}{\bf{Num. of Incon. Behaviors}} \\ \cline{3-6}
\bf{Scene} & & \ang{10} & \ang{20} & \ang{30} & \ang{40} \\ \hline \hline

\multirow{ 3}{*}{Snowy} & Autumn & 11635 & 11602 & 11388 & 10239\\ 
                       & Chauffeur & 4839 & 2105 & 1093 & 653\\
                       & Rwightman & 334 & 115 & 45 & 14\\ \hline

\multirow{ 3}{*}{Rainy} & Autumn & 5279 & 5279 & 5279 & 5279\\ 
                       & Chauffeur & 710 & 175 & 94 & 71\\
                       & Rwightman & 656 & 92 & 23 & 0\\ \hline
\end{tabular}
\end{table}

\section{Related work}
\label{sec5}

\noindent\textbf{Testing and verification of DNN-based autonomous driving systems.} Different from traditional testing practices for DNN models~\cite{witten2016data,madrigal2017inside}, a recent set of approaches (such as DeepXplore~\cite{pei2017deepxplore} and DeepTest~\cite{DeepTest}) utilize differential and metamorphic testing algorithms for identifying inputs that trigger inconsistencies among different DNN models, or among the original and transformed driving scenes. Although such approaches have successfully found various autonomous driving system issues, there still lack approaches that can test DNN-based autonomous driving system with realistic synthesized driving scenes.\Comment{ DeepXplore developed the concept of neuron coverage for systematically measuring the amount of tested internal logic of a DNN. DeepTest further used the neuron coverage metric for generating guided tests to find erroneous driving behaviors in a single DNN through leveraging metamorphic testing. }

% \noindent\textbf{GAN-based Techniques.} GAN-based domain adaption has been recently shown to be effective in unsupervised image-to-image translation~\cite{zhu2017unpaired,kim2017learning,yi2017dualgan,liu2017unsupervised}. CycleGan~\cite{zhu2017unpaired}, DiscoGAN~\cite{kim2017learning} and DualGan~\cite{yi2017dualgan} propose the same idea that image-to-image translation should satisfy the cycle consistency, where an image from Domain A should be identical when it is translated to Domain B and then translated back to Domain A. The experiments show that this extra constraint can make the translated images more realistic. UNIT~\cite{liu2017unsupervised} further assumes that the representations of two domains may be projected to the same vector space (shared latent space), and is constructed based on VAEs and GANs. Specifically, they also apply cycle consistency to the GAN model to regularize the translation.

\noindent\textbf{GAN-based virtual/real scene adaption.} GAN-based domain adaption has been recently shown to be effective in virtual-to-real and real-to-virtual scene adaption~\cite{yang2018unsupervised,li2018semantic}. DU-drive~\cite{yang2018unsupervised} proposes an unsupervised real to virtual domain unification framework for end-to-end driving. Their key insight is the raw image may contain nuisance details which are not related to the prediction of steering angles, and a corresponding virtual scene can ignore these details and also address the domain shift problem. SG-GAN~\cite{li2018semantic} is designed to automatically transfer the scene annotation in virtual-world to facilitate real-world visual tasks. In that work, a semantic-aware discriminator is proposed for validating the fidelity of rendered image w.r.t each semantic region. 

% unsupervised image-to-image translation~\cite{zhu2017unpaired,kim2017learning,yi2017dualgan,liu2017unsupervised}. CycleGan~\cite{zhu2017unpaired}, DiscoGAN~\cite{kim2017learning} and DualGan~\cite{yi2017dualgan} propose the same idea that image-to-image translation should satisfy the cycle consistency, where an image from Domain A should be identical when it is translated to Domain B and then translated back to Domain A. The experiments show that this extra constraint can make the translated images more realistic. UNIT~\cite{liu2017unsupervised} further assumes that the representations of two domains may be projected to the same vector space (shared latent space), and is constructed based on VAEs and GANs. Specifically, they also apply cycle consistency to the GAN model to regularize the translation.

% However, most of such works, e.g., DeepTest, perform rather simple synthetic image transformation, such as adding simple weather effect filters, resulting in unrealistic and superficial driving scenes. To fundamentally complement DeepTest, we design \OurApproach{} to leverage GAN-based techniques for automatically delivering authentic driving scenes with various weather conditions which are rather difficult to be collected manually.\lll{Mengshi can you modify this paragraph since you are familiar with their work?}

\noindent\textbf{Metamorphic testing.} Metamorphic testing is a classical software testing method that identify software bugs~\cite{zhou2004metamorphic,chen1998metamorphic,segura2016survey}. Its core idea is to detect violations of domain-specific metamorphic relations defined across outputs from multiple runs of the  program with different inputs. Metamorphic testing has been applied for testing machine learning classifiers~\cite{murphy2008properties,xie2009application,xie2011testing}. In this paper, \OurApproach{} develops a specific GAN-based metamorphic testing module for DNN-based autonomous systems, where the metamorphic
relations are defined such that regardless of how the driving
scenes are synthesized to cope with weather conditions, the
driving behaviors are expected to be consistent with those under
the corresponding original driving scenes. 

\section{Conclusion}
\label{sec6}

In this paper, we propose \OurApproach{}, an unsupervised GAN-based approach to synthesize authentic driving scenes with various weather conditions to test DNN-based autonomous driving systems. In principle, \OurApproach{} applies the metamorphic testing methodology to detect the inconsistent autonomous driving behaviors across different driving scenes. The experimental results on three real-world Udacity autonomous driving models indicate that \OurApproach{} can  successfully detect thousands of inconsistent behaviors. Furthermore, our results also show that \OurApproach{} can be promising in measuring the robustness of autonomous driving systems.  Currently, \OurApproach{} only supports two weather conditions, we plan to support more weather conditions to fully test autonomous driving systems under various conditions in the near future.

\Comment{In the future, we would refine the GAN-based techniques such that synthesizing authentic driving scenes could more efficient. We would also like to add more effects to \OurApproach{} such that it could be more robust to handle more road conditions. }

\footnotesize
\balance
\bibliographystyle{plain}
\bibliography{section_ref}

\begin{thebibliography}{10}

\bibitem{UdRank}
Final leaderboard of udacity challenge 2.
\newblock
  \url{https://github.com/udacity/self-driving-car/tree/master/challenges/challenge-2}.

\bibitem{Autumn}
Steering angle model: Autumn.
\newblock
  \url{https://github.com/udacity/self-driving-car/tree/master/steering-models/evaluation}.

\bibitem{Chauffeur}
Steering angle model: Chauffeur.
\newblock
  \url{https://github.com/udacity/self-driving-car/tree/master/steering-models/community-models/chauffeur}.

\bibitem{Rwightman}
Steering angle model: Rwightman.
\newblock
  \url{https://github.com/udacity/self-driving-car/tree/master/steering-models/evaluation}.

\bibitem{TeslaAutopilot}
Tesla autopilot system.
\newblock \url{https://www.tesla.com/autopilot}.

\bibitem{TeslaCrash}
Tesla model s crash.
\newblock
  \url{https://www.teslarati.com/tesla-model-s-firetruck-crash-details/}.

\bibitem{UdacityDataset}
Udacity challenge 2 dataset.
\newblock
  \url{https://github.com/udacity/self-driving-car/tree/master/datasets}.

\bibitem{ModelCombo}
Udacity pre-trained models.
\newblock
  \url{https://github.com/udacity/self-driving-car/tree/master/steering-models/evaluation}.

\bibitem{Udacity}
Udacity self driving car.
\newblock \url{https://github.com/udacity/self-driving-car}.

\bibitem{YoutubeSnow}
Youtube video: Driving on snow - greenville, nc 1-4-2018 at 7:00am.
\newblock \url{https://www.youtube.com/watch?v=ps56tnQG8V0}.

\bibitem{YoutubeRain}
Youtube video: Rain on a car roof - 1 hour - asmr.
\newblock \url{https://www.youtube.com/watch?v=O88fXBx-Qdg}.

\bibitem{ammann2016introduction}
Paul Ammann and Jeff Offutt.
\newblock {\em Introduction to software testing}.
\newblock Cambridge University Press, 2016.

\bibitem{bojarski2016end}
Mariusz Bojarski, Davide Del~Testa, Daniel Dworakowski, Bernhard Firner, Beat
  Flepp, Prasoon Goyal, Lawrence~D Jackel, Mathew Monfort, Urs Muller, Jiakai
  Zhang, et~al.
\newblock End to end learning for self-driving cars.
\newblock {\em arXiv preprint arXiv:1604.07316}, 2016.

\bibitem{chen1998metamorphic}
Tsong~Y Chen, Shing~C Cheung, and Siu~Ming Yiu.
\newblock Metamorphic testing: a new approach for generating next test cases.
\newblock Technical report, Technical Report HKUST-CS98-01, Department of
  Computer Science, Hong Kong University of Science and Technology, Hong Kong,
  1998.

\bibitem{goodfellow2014generative}
Ian Goodfellow, Jean Pouget-Abadie, Mehdi Mirza, Bing Xu, David Warde-Farley,
  Sherjil Ozair, Aaron Courville, and Yoshua Bengio.
\newblock Generative adversarial nets.
\newblock In {\em Advances in neural information processing systems}, pages
  2672--2680, 2014.

\bibitem{kingma2013auto}
Diederik~P Kingma and Max Welling.
\newblock Auto-encoding variational bayes.
\newblock {\em arXiv preprint arXiv:1312.6114}, 2013.

\bibitem{krizhevsky2012imagenet}
Alex Krizhevsky, Ilya Sutskever, and Geoffrey~E Hinton.
\newblock Imagenet classification with deep convolutional neural networks.
\newblock In {\em Advances in neural information processing systems}, pages
  1097--1105, 2012.

\bibitem{li2018semantic}
Peilun Li, Xiaodan Liang, Daoyuan Jia, and Eric~P Xing.
\newblock Semantic-aware grad-gan for virtual-to-real urban scene adaption.
\newblock {\em arXiv preprint arXiv:1801.01726}, 2018.

\bibitem{liu2017unsupervised}
Ming-Yu Liu, Thomas Breuel, and Jan Kautz.
\newblock Unsupervised image-to-image translation networks.
\newblock In {\em Advances in Neural Information Processing Systems}, pages
  700--708, 2017.

\bibitem{madrigal2017inside}
Alexis~C Madrigal.
\newblock Inside waymo’s secret world for training self-driving cars.
\newblock {\em The Atlantic}, 2017.

\bibitem{mckeeman1998differential}
William~M McKeeman.
\newblock Differential testing for software.
\newblock {\em Digital Technical Journal}, 10(1):100--107, 1998.

\bibitem{murphy2008properties}
Christian Murphy, Gail~E Kaiser, Lifeng Hu, and Leon Wu.
\newblock Properties of machine learning applications for use in metamorphic
  testing.
\newblock In {\em SEKE}, volume~8, pages 867--872, 2008.

\bibitem{pei2017deepxplore}
Kexin Pei, Yinzhi Cao, Junfeng Yang, and Suman Jana.
\newblock Deepxplore: Automated whitebox testing of deep learning systems.
\newblock In {\em Proceedings of the 26th Symposium on Operating Systems
  Principles}, pages 1--18. ACM, 2017.

\bibitem{Pomerlea}
Dean~A. Pomerleau.
\newblock Advances in neural information processing systems 1.
\newblock chapter ALVINN: An Autonomous Land Vehicle in a Neural Network, pages
  305--313. Morgan Kaufmann Publishers Inc., San Francisco, CA, USA, 1989.

\bibitem{sak2014long}
Ha{\c{s}}im Sak, Andrew Senior, and Fran{\c{c}}oise Beaufays.
\newblock Long short-term memory recurrent neural network architectures for
  large scale acoustic modeling.
\newblock In {\em Fifteenth annual conference of the international speech
  communication association}, 2014.

\bibitem{metamorphic}
S.~Segura, G.~Fraser, A.~B. Sanchez, and A.~Ruiz-Cortés.
\newblock A survey on metamorphic testing.
\newblock {\em IEEE Transactions on Software Engineering}, 42(9):805--824, Sept
  2016.

\bibitem{segura2016survey}
Sergio Segura, Gordon Fraser, Ana~B Sanchez, and Antonio Ruiz-Cort{\'e}s.
\newblock A survey on metamorphic testing.
\newblock {\em IEEE Transactions on software engineering}, 42(9):805--824,
  2016.

\bibitem{DeepTest}
Yuchi Tian, Kexin Pei, Suman Jana, and Baishakhi Ray.
\newblock Deeptest: Automated testing of deep-neural-network-driven autonomous
  cars.
\newblock In {\em Proceedings of the 40th International Conference on Software
  Engineering, Gothenburg, Sweden, May 27 - June 3, 2018}, ICSE 2018, 2018.

\bibitem{witten2016data}
Ian~H Witten, Eibe Frank, Mark~A Hall, and Christopher~J Pal.
\newblock {\em Data Mining: Practical machine learning tools and techniques}.
\newblock Morgan Kaufmann, 2016.

\bibitem{xie2009application}
Xiaoyuan Xie, Joshua Ho, Christian Murphy, Gail Kaiser, Baowen Xu, and
  Tsong~Yueh Chen.
\newblock Application of metamorphic testing to supervised classifiers.
\newblock In {\em Quality Software, 2009. QSIC'09. 9th International Conference
  on}, pages 135--144. IEEE, 2009.

\bibitem{xie2011testing}
Xiaoyuan Xie, Joshua~WK Ho, Christian Murphy, Gail Kaiser, Baowen Xu, and
  Tsong~Yueh Chen.
\newblock Testing and validating machine learning classifiers by metamorphic
  testing.
\newblock {\em Journal of Systems and Software}, 84(4):544--558, 2011.

\bibitem{yang2018unsupervised}
Luona Yang, Xiaodan Liang, and Eric Xing.
\newblock Unsupervised real-to-virtual domain unification for end-to-end
  highway driving.
\newblock {\em arXiv preprint arXiv:1801.03458}, 2018.

\bibitem{zhou2004metamorphic}
Zhi~Quan Zhou, DH~Huang, TH~Tse, Zongyuan Yang, Haitao Huang, and TY~Chen.
\newblock Metamorphic testing and its applications.
\newblock In {\em Proceedings of the 8th International Symposium on Future
  Software Technology (ISFST 2004)}, pages 346--351, 2004.

\end{thebibliography}
\end{document}